\documentclass[12pt,emulateapj]{emulateapj}
\usepackage{lscape}

\def \kms{\mbox{kms$^{-1}$}}
\def \arcsec{\mbox{$^{\prime\prime}$}}

\def \apjs{ApJS}

\def \apjl{ApJL}

\def \aap{A\&A}

\newcommand{\Msun}{M$_{\odot}$}

\newcommand{\cmc}{${\rm cm}^{-3}$}

\newcommand{\cloud}{{\rm G0.253+0.016}}

\begin{document}

\shorttitle{$\cloud$: a molecular cloud progenitor of an Arches-like cluster}

\shortauthors{S. N. Longmore et al.}

\title{$\cloud$: a molecular cloud progenitor of an Arches-like cluster}

\author{Steven N. Longmore$^{1,2}$, Jill Rathborne$^3$, Nate Bastian$^4$,
  Joao Alves$^5$, Joana Ascenso$^1$, John Bally$^7$, Leonardo
  Testi$^1$, Andy Longmore$^6$, Cara Battersby$^7$, Eli Bressert$^{1,8}$,
  Cormac Purcell$^{9,10}$, Andrew Walsh$^{11}$, James Jackson$^{12}$, Jonathan
  Foster$^{12}$, Sergio Molinari$^{13}$, Stefan Meingast$^5$,
  A. Amorim$^{14}$, J. Lima$^{14}$, R. Marques$^{14}$,
  A. Moitinho$^{14}$, J. Pinhao$^{15}$, J. Rebordao$^{16}$,
  F. D. Santos$^{14}$}

\affil{$^1$European Southern Observatory, Karl-Schwarzschild-Str. 2,
  85748 Garching bei Munchen, Germany\\
$^2$Harvard-Smithsonian Center for Astrophysics, 60 Garden Street, Cambridge, MA 02138, USA\\
$^3$CSIRO Astronomy and Space Science, Epping, Sydney, Australia\\
$^4$Excellence Cluster Universe, Boltzmannstr. 2, 85748 Garching, Germany\\
$^5$University of Vienna, T\"urkenschanzstrasse 17, 1180 Vienna, Austria\\
$^6$UK Astronomy Technology Centre, Edinburgh, UK\\
$^7$Center for Astrophysics and Space Astronomy, University of Colorado, UCB 389, Boulder, CO 80309\\
$^8$School of Physics, University of Exeter, Stocker Road, Exeter EX4 4QL, UK \\
$^9$Department of Astronomy, University of Leeds, UK\\
$^{10}$Sydney Institute for Astronomy (SiFA), School of Physics, The University of Sydney, NSW 2006, Australia\\
$^{11}$Department of Astronomy, James Cook University, Townesville, Australia\\
$^{12}$Institute for Astrophysical Research, Boston University, Boston, MA 02215, USA\\
$^{13}$INAF, Rome, Italy\\
$^{14}$SIM, Faculdade de Ci\^encias da Universidade de Lisboa, Ed. C8. Campo Grande 1749-016 Lisbon, Portugal \\
$^{15}$LIP-Coimbra, Department of Physics, University of Coimbra, 3004-516 Coimbra, Portugal\\
$^{16}$CAAUL-FCUL, Ed. D. Est. Paco Lumiar 22, 1649-038 Lisbon, Portugal\\
}

\email{slongmor@eso.org}

\begin{abstract}

Young massive clusters (YMCs) with stellar masses of
$10^4-10^5$\,M$_\odot$ and core stellar densities of $10^4-10^5$ stars
per cubic pc are thought to be the `missing link' between open
clusters and extreme extragalactic super star clusters and globular
clusters. As such, studying the initial conditions of YMCs offers an
opportunity to test cluster formation models across the full cluster
mass range. $\cloud$ is an excellent candidate YMC progenitor.  We
make use of existing multi-wavelength data including recently
available far-IR continuum (Herschel/Hi-GAL) and mm spectral line
(HOPS and MALT90) data and present new, deep, multiple-filter, near-IR
(VLT/NACO) observations to study $\cloud$. These data show $\cloud$ is
a high mass ($1.3\times10^5\,$M$_\odot$), low temperature (T$_{\rm
  dust}\sim$20\,K), high volume and column density
(n~$\sim8\times10^4$\,cm$^{-3}$;
N$_{H_2}$~$\sim4\times10^{23}$\,cm$^{-2}$) molecular clump which is
close to virial equilibrium (M$_{\rm dust}~\sim~M_{\rm virial}$) so is
likely to be gravitationally-bound. It is almost devoid of star
formation and, thus, has exactly the properties expected for the
initial conditions of a clump that may form an Arches-like massive
cluster. We compare the properties of $\cloud$ to typical Galactic
cluster-forming molecular clumps and find it is extreme, and possibly
unique in the Galaxy. This uniqueness makes detailed studies of
$\cloud$ extremely important for testing massive cluster formation
models.

\end{abstract}


\section{INTRODUCTION}

One of the ground breaking discoveries of the Hubble Space Telescope
was that massive stellar clusters, with properties that rival (or even
exceed) those found in globular clusters in terms of mass and stellar
density, are still forming in the universe today
\citep{holtzman1992}. These young massive clusters (YMCs) can be
broadly defined as stellar clusters more massive than
10$^4$\,M$_\odot$ and with ages that are $<$100\,Myr but that also
exceed the current crossing time by a factor of a few, so are
gravitationally-bound \citep[see e.g.][]{ashman_zepf2001,
  portegieszwart2010} . The discovery of YMCs in the Galaxy has
allowed a detailed study of their structural and stellar properties
\citep[e.g. Arches, Quintuplet, Westerlund 1 and 2, RSCG1, 2 and 3,
  GLIMPSE-CO1,
  NGC~3603:][]{figer1999,clark2005,ascenso2007,figer2006,davies2011,beccari2010}.
Like their extragalactic counterparts, these Galactic YMCs have
stellar masses of $10^4-10^5$\,M$_\odot$ and core stellar densities of
$10^4-10^5$ stars per cubic pc.  The stellar masses and densities are
orders of magnitude larger than typical open clusters and comparable
to those in globular clusters, super star clusters seen in
merging/starburst galaxies and those inferred from observations of
massive clusters forming at the epoch of peak star formation density
($z\sim2-3$) \citep{whitmore2000, johnson_kobulnicky2003,
  swinbank2010, danielson2011}. As such, YMCs are potentially
local-universe-analogs of these massive extragalactic and high-$z$
stellar clusters. Within our own Galaxy, observations show a smooth
continuum in the properties of young clusters, suggesting that YMCs
merely represent the extreme end of the open cluster (OC) distribution
function. Thus, understanding how YMCs form may provide a direct
handle on i) globular and super star cluster formation, ii) cluster
formation at the epoch of peak star formation density ($z\sim2-3$),
iii) the most extreme conditions for star formation in the local
universe, and iv) whether or not there are different formation modes
across the cluster mass range.

Despite their importance, the formation process of YMCs is not well
constrained observationally due to the lack of known potential gas
clouds as progenitors. As a result it has not been possible to study
the initial conditions of the molecular gas from which YMCs form
\citep[see e.g.][]{santangelo2009}. Progress requires first finding
and then deriving the properties of the most massive molecular gas
clouds which are on the verge of forming YMCs. Such YMC progenitor
clouds are expected to have a high mass ($\sim$10$^5$\,M$_\odot$) and
small radii (a few pc) with high volume and column densities
($\gtrsim$10$^4$\,cm$^{-3}$; $\gtrsim$10$^{23}$\,cm$^{-2}$). Before
the onset of star formation, the gas is also expected to be very cold
($\sim10-20$\,K). In order to resolve the internal structure of
objects with radii $\sim$pc requires limiting the search to gas clouds
within our own Galaxy.

Compared to the objects found in recent surveys of dense
cluster-forming molecular clumps, one object, $\cloud$, stands out as
extreme \citep[e.g.][]{sridharan2002,hill2005, beuther2002,
  schuller2009, bally2010, aguirre2011, molinari2010,
  rathborne2006irdc, simon2006irdcs, simon2006irdc2, pillai2006,
  peretto_fuller2009, jackson2006grs, rathborne2009, longmore2007,
  purcell2006, purcell2009,
  walsh2008,walsh2011,jackson2011}. Identified as an infrared dark
cloud (IRDC), it is clearly seen as a prominent extinction feature in
mid-IR images (see Figure~\ref{fig:cont_image}). \citet{molinari2011}
place $\cloud$ in the ``elliptical ring'' feature orbiting the
Galactic center, putting it $60-100$\,pc in front of the Galactic
center, at a distance of $\sim$8.4\,kpc \citep{reid2009}.  With a dust
temperature of $\sim$20~K, volume density of $>$10$^{4}$\,\cmc, dust
mass of $\sim$10$^{5}$\,\Msun, and almost no signs of star-formation,
$\cloud$, has exactly the properties expected for the precursor to a
massive, Arches-like cluster \citep{lis1994,lis1998,lis2001,
  bally2010}.

Despite having the expected global properties for a YMC precursor, the
dynamical state and fate of future star formation in $\cloud$ remains
uncertain. While the detection of a weak water maser suggests there
may be a small number of low-mass stars forming \citep{lis1994}, it is
unclear whether the gas is globally bound and collapsing (and
therefore going to form a massive cluster) or unbound and transient
\citep{lis1998}. This ambiguity is complicated by the extreme
environmental conditions within the Galactic center region
(e.g. crowding, intense radiation fields, large magnetic fields, tidal
shear). Given its importance as a potential precursor to a YMC, we
make use of higher resolution and sensitivity data to study $\cloud$
in detail and attempt to ascertain its dynamical state.

\section{Survey data and new observations}
\label{sub:obs}

\subsection{Survey data}
\label{sub:survey_data}

We make use of multi-wavelength archival data from the United Kingdom
Infrared Digital Sky Survey \citep[UKIDSS:][data release
  7]{lawrence2007}, the Spitzer Galactic Legacy Infrared Mid-Plane
Survey Extraordinaire \citep[GLIMPSE:][]{benjamin2003}, the
Submillimetre Common-User Bolometer Array (SCUBA) on the James Clerk
Maxwell Telescope (JCMT) \citep{difrancesco2008} and the methanol
multi-beam survey \citep[MMB:][]{caswell2010}. We also make use of
recently available data from the Herschel Infrared Galactic Plane
Survey \citep[HiGAL:][]{molinari2010PASP, molinari2011}, the H$_2$O
southern Galactic plane survey \citep[HOPS:][]{walsh2008, walsh2011}
and the Millimetre Astronomy Legacy Team 90\,GHz Survey
\citep[MALT90:][]{jackson2011,foster2011}. Details of the surveys
including observational parameters and references outlining the data
acquisition/reduction procedures can be found in
Table~\ref{tab:obs_summary}.

\begin{table*}
  \caption{Summary of observational survey data used in this work}
{\scriptsize
  \begin{tabular}{ccccccccccc}
    \hline
    Survey/Archive  & Telescope(s)   & $\lambda$                      & continuum/line & $\theta$         & Reference(s)          \\ \hline
    UKIDSS          & UKIRT          & 1.2, 1.6, 2.2\,$\mu$m          & continuum      & $<$1$\arcsec$    & \citet{lawrence2007}  \\
    GLIMPSE         & Spitzer        & 3.6, 4.5, 5.6, 8.0\,$\mu$m     & continuum      & 2$\arcsec$       & \citet{benjamin2003}  \\
    HiGAL           & Herschel       & 70, 160, 250, 350, 500\,$\mu$m & continuum      & 5$-$36$\arcsec$  & \citet{molinari2010PASP,molinari2011} \\
    SCUBA Archive   & JCMT           & 450, 850\,$\mu$m               & continuum      & 8$\arcsec$       & \citep{difrancesco2008} \\
    MALT90          & Mopra          & 3\,mm                          & line           & 35$\arcsec$      & \citep{jackson2011,foster2011} \\
    HOPS            & Mopra          & 12\,mm                         & line           & 2$\arcmin$       & \citep{walsh2008, walsh2011} \\
    MMB             & Parkes \& ATCA & 3\,cm                          & line           & $\sim$1$\arcsec$ & \citep{caswell2010} \\ \hline\hline

  \end{tabular}
}
  \label{tab:obs_summary}

\end{table*}

\subsection{VLT/NACO Observations}
\label{sub:vlt_obs}

Deep, near-IR data were obtained with the Nasmyth Adaptive Optics
System Near-Infrared Imager and Spectrograph (NAOS-CONICA, or NACO) on
the Very Large Telescope (VLT) on May 5th 2011. A field of view of $1'
\times 1'$ centered on $17^h42^m58^s$, $-28^\circ42'23"$ (J2000) was
imaged in $J$, $H$ and $K_s$. Although offset from the nominal center
of $\cloud$, this position was chosen to be centered on a foreground
star which was bright enough to use the Natural Guide Star adaptive
optics system guiding with camera S54 and dichroic N20C80. After
flatfielding and sky subtraction using IRAF\footnote{IRAF is
  distributed by the National Optical Astronomy Observatories, which
  are operated by the Association of Universities for Research in
  Astronomy, Inc., under cooperative agreement with the National
  Science Foundation.}  standard tools, the six dithered exposures of
10s each in each filter were shifted and combined into the final $J$,
$H$ and $K_s$ images. The source extraction and PSF photometry were
done with IRAF DAOPHOT using standard techniques and error cuts. We
used Topcat\footnote{\citet{taylor2005}:
  http://www.star.bris.ac.uk/$\sim$mbt/topcat/} to match all the stars
towards $\cloud$ observed by VLT/NACO with the UKIDSS data (see
$\S$~\ref{sub:near_ir_extinction}). Differences in filter wavelengths
were corrected to first order using the principles described in
\citet{stead_hoare_2009} and Appendix A of
\citet{schodel2010}. Remaining uncertainties in color effects and
from the statistical scatter mean the error in calibration is no more
than $\pm0.1$\,mag at each wavelength. The final NACO catalog contains
358, 1001 and 1830 sources, detected to the $5\sigma$ level up to
21.3, 20.5 and 19.8 mag in $J$, $H$ and $K_s$, respectively. Despite
the error cuts, up to 2\% of the sources in the catalog in each filter
are still spurious detections.

\section{Results}
\label{sec:results}

\begin{figure*}
  \begin{center}
    \begin{tabular}{ccc}
    \includegraphics[width=0.3\textwidth,clip=true]{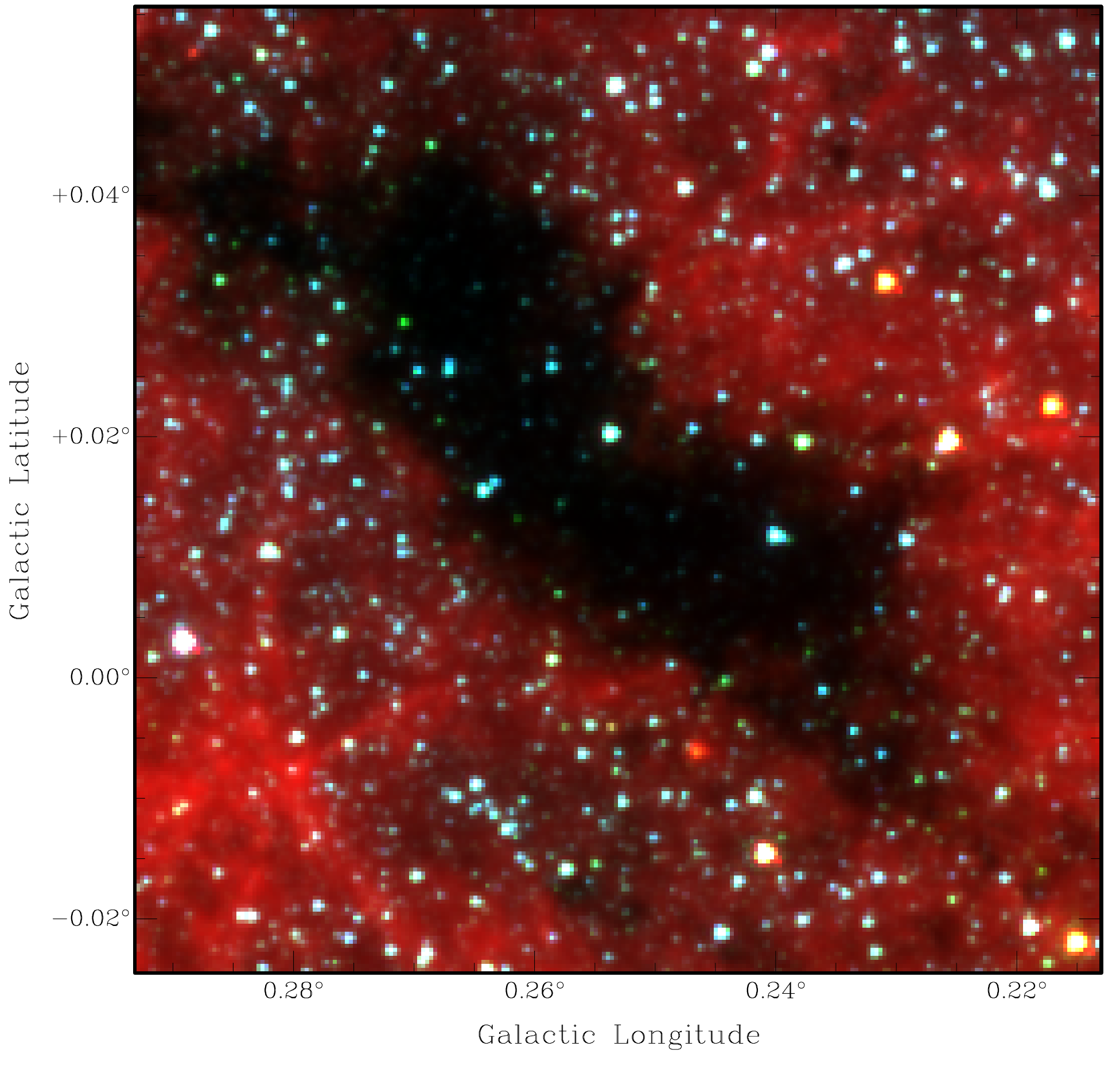} &
    \includegraphics[width=0.305\textwidth,clip=true]{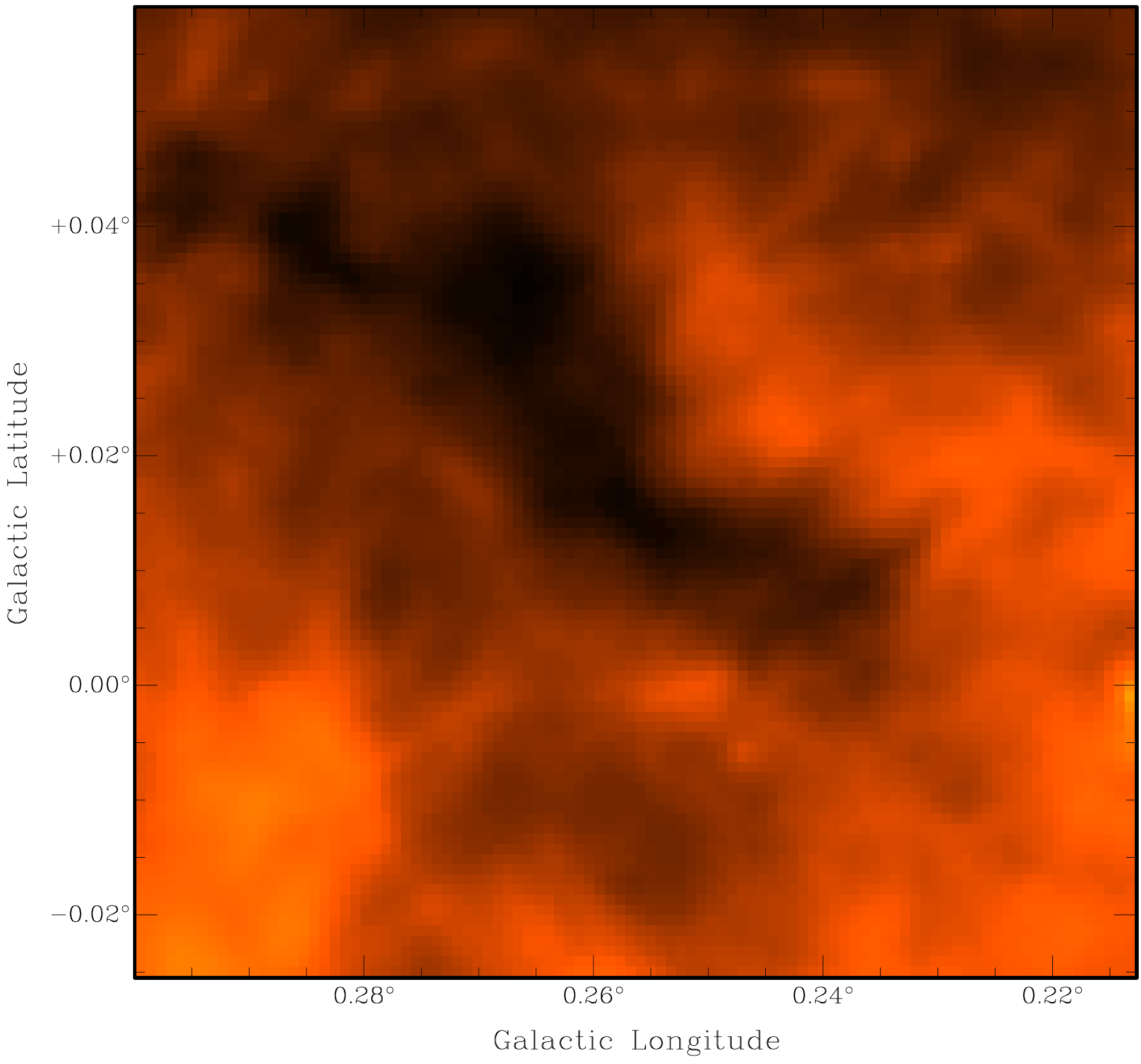}&
    \includegraphics[width=0.3\textwidth,clip=true]{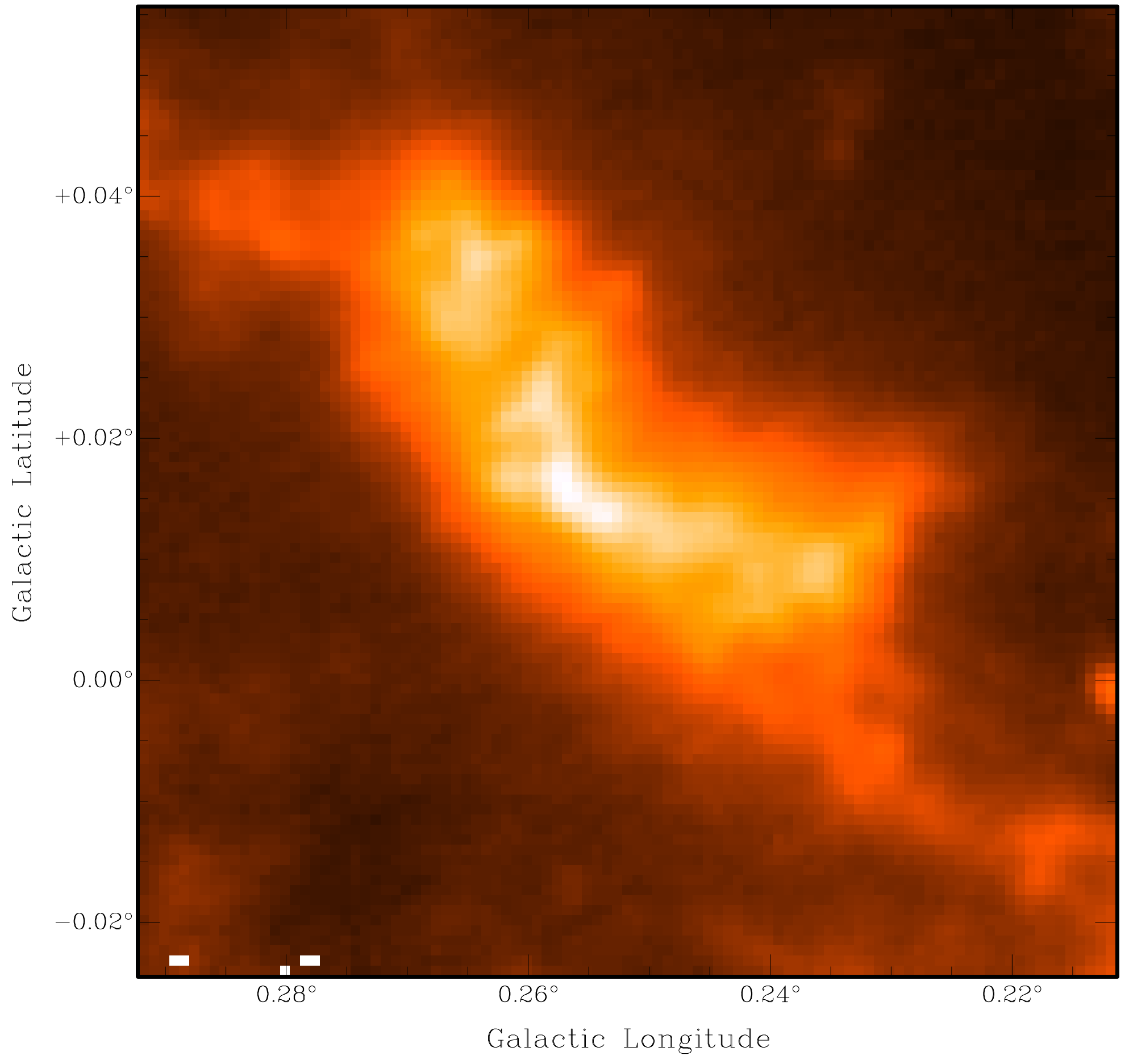} \\
    \end{tabular}
    \caption{Continuum images toward \cloud. The images are 14\,pc on
      a side for the distance of 8.4\,kpc. {\it Left to right:}
      GLIMPSE \citep{benjamin2003} three color (3.6, 4.5 and
      8\,$\mu$m), Herschel 70\,$\mu$m \citep{molinari2011}, SCUBA/JCMT
      450\,$\mu$m \citep{difrancesco2008}. The cloud \cloud, is seen
      as an extinction feature in the mid-IR to far-IR but is a strong
      emitter in the sub-mm/mm. As such it must be both cold and dense
      and sit in front of the majority of the diffuse Galactic mid-IR
      background emission.}
    \label{fig:cont_image}
\end{center} 
\end{figure*}

Figure~\ref{fig:cont_image} shows continuum images of $\cloud$ at
3.6$-$8.0$\mu$m, 70$\mu$m and 450$\mu$m. $\cloud$ is elongated with an
aspect ratio of approximately 3:1 and is seen as an extinction feature
from 3.6 to 70$\mu$m with no obvious embedded emission sources. As
such it must be both cold and dense and sit in front of the majority
of the diffuse Galactic mid-IR background emission. At wavelengths
$\ge$170$\mu$m, bright emission is seen across $\cloud$
\citep{molinari2011} with similar morphology to the 450$\mu$m emission
shown in Figure~\ref{fig:cont_image}. The lack of embedded emission
sources at wavelengths up to 70$\mu$m combined with results from
recent maser surveys which show no 6.7\,GHz CH$_3$OH masers
\citep{caswell2010} or new 22\,GHz H$_2$O maser detections
\citep{walsh2011} strongly reinforces the paucity of active star
formation within the cloud.

\subsection{Near-IR extinction}
\label{sub:near_ir_extinction}

From continuum observations alone it is not possible to determine
whether the emission from $\cloud$ arises from a single physical
entity or from multiple clouds along the same line of
sight. Extinction measurements can be used to distinguish between
these two scenarios. Additionally, given the confusion and the
difficulty deriving kinematic distances towards the Galactic center,
extinction measurements also offer an independent distance
determination.

\begin{figure*}
 \begin{center}
   \begin{tabular}{cc}
 	\includegraphics[width=0.4\textwidth,angle=0]{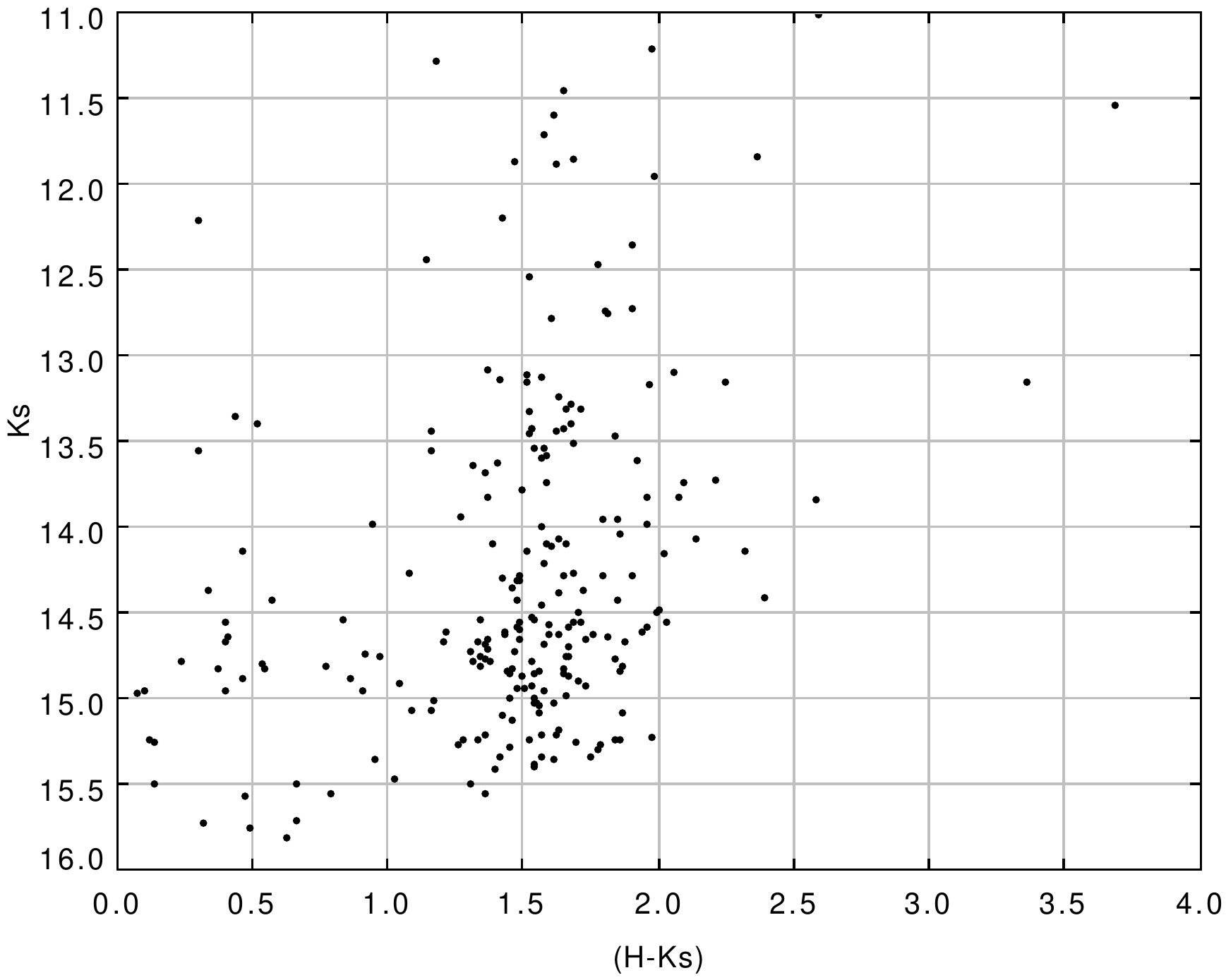} &
        \includegraphics[width=0.4\textwidth,angle=0]{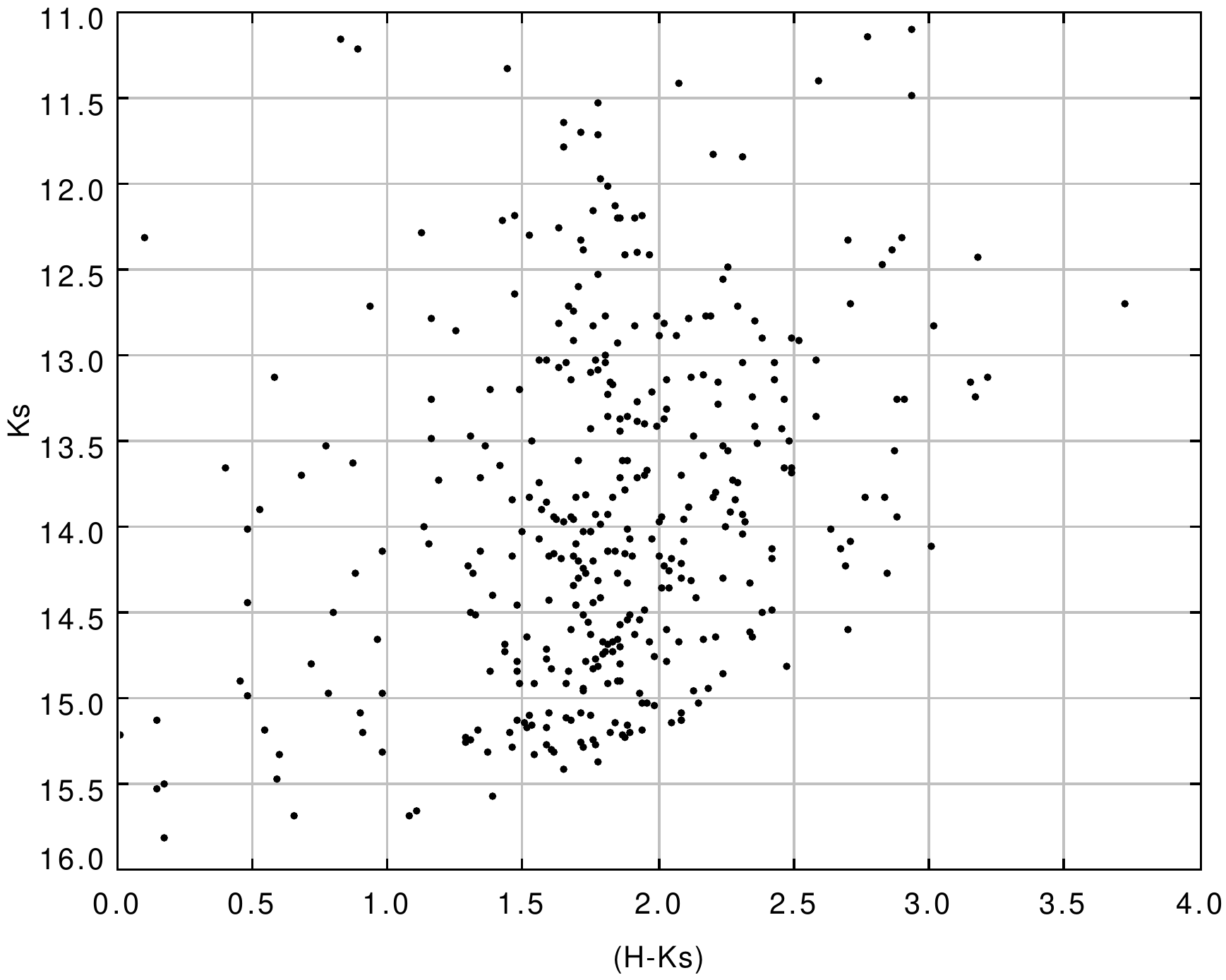}\\
 	\includegraphics[width=0.4\textwidth,angle=0]{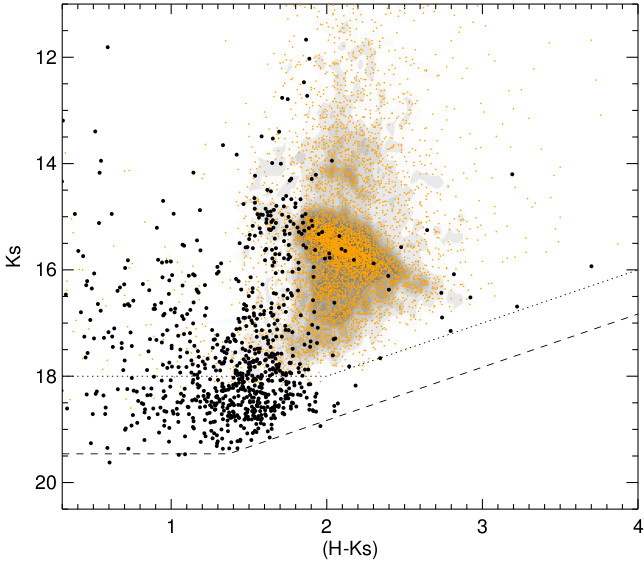} &
        \includegraphics[width=0.4\textwidth,angle=0]{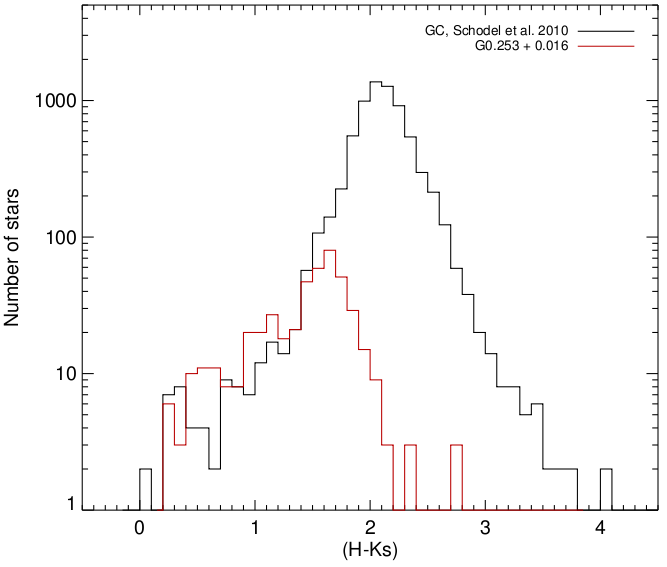} \\
   \end{tabular}
 	\caption{\small{[Top panel] $K_s$ {\it vs.} $(H-K_s)$
            color-magnitude diagrams for UKIDSS sources within a
            circle of radius 45$\arcsec$ centered on $\cloud$ [left]
            and an offset position at $l = 0.32^\circ$, $b =
            -0.02^\circ$ [right]. Although the UKIDSS data are
            confusion-limited in sensitivity and not as deep as the
            NACO data (bottom panel) a clear decrease in source counts
            towards $\cloud$ is seen at $H-K_s \sim 1.7 \pm 0.2$\,mag.
            [Bottom left panel] $K_s$ {\it vs.}  $(H-K_s)$
            color-magnitude diagram. The black dots show sources
            detected in the VLT/NACO data towards $\cloud$ (see
            $\S$~\ref{sub:vlt_obs}~\&~\ref{sub:near_ir_extinction}). The
            yellow dots and grey scale show the VLT/NACO sources and
            number density, respectively, detected towards the
            Galactic center from \citet{schodel2010}. Both VLT/NACO
            observations have the same $1' \times 1'$ field of view
            but the $\cloud$ observations are $\sim$2\,mag deeper (as
            illustrated by the dotted and dashed lines showing the
            approximate 5$\sigma$ sensitivity limits of the
            \citet{schodel2010} and $\cloud$ datasets,
            respectively). In the \citet{schodel2010} data, the red
            giant branch bump feature continues to
            $(H-K_s)$~$\sim$~2.4 mag and is still visible at 2.6 mag,
            where it becomes strongly effected by completeness
            limits. Despite being $\sim$2\,mag deeper and looking
            along similar lines of sight (and hence similar foreground
            extinction and expected stellar populations), the red
            giant branch bump feature is sharply truncated in the
            $\cloud$ observations.  [Bottom right panel] Histogram of
            the $(H-K_s)$ colors for the \citet{schodel2010} catalog
            [black], and for the NACO sources in $\cloud$ brighter
            than the \citet{schodel2010} detection limit [red]. Both
            distributions are similar for $(H-K_s)$~$\leq$~1.7 but the
            number of $\cloud$ sources shows a sharp cutoff at redder
            colors, despite the observations being $\sim$2\,mag
            deeper. This is clear evidence that we have detected the
            extinction from $\cloud$, and that this is a single entity
            as opposed to multiple clouds along the line of sight. The
            distance determined from these data is consistent with
            $\cloud$ being in close proximity to the Galactic center
            ($\S$~\ref{sub:near_ir_extinction}).}}
    	 \label{fig:extinction}
 \end{center} 
 \end{figure*}

We used data from UKIDSS, specifically the Galactic Plane Survey
\citep[GPS:][]{lucas2008}, and deep VLT/NACO observations (see
$\S$~\ref{sub:vlt_obs}) to investigate the near-IR extinction towards
$\cloud$. We downloaded the infrared sources in a 15$\arcmin$ x
15$\arcmin$ region centered on $\cloud$ from the UKIDSS data base,
which allowed for a direct and coherent comparison of the cloud and
off-cloud properties. The comparison of on- and off-source $K_s$ {\it
  vs.} $(H-K_s)$ color-magnitude diagrams (CMDs) showed that there are
far fewer very red stars in the direction of $\cloud$ than in any of
the off-source diagrams, with a clear and maintained deficit in the
number of red stars well established by $(H-K_s) = 1.7 \pm 0.2$\,mag
(see top panel of Figure~\ref{fig:extinction})\footnote{ It should be
  noted that across the UKIDSS field examined there is also
  considerable variation in the foreground extinction, not surprising
  in a field so close to the Galactic Centre direction.  }. This is
equally clearly seen when comparing the NACO data for $\cloud$ with
similar field of view, albeit shallower, NACO observations toward the
Galactic Centre by \citet{schodel2010}: Figure~\ref{fig:extinction}
shows the $K_s$ {\it vs.} $(H-K_s)$ CMD, and the histogram of
$(H-K_s)$ colors for that Galactic Centre field and for
$\cloud$. Toward the Galactic center, the red giant branch bump
feature is clearly seen extending to $(H-K_s)$~$\sim$~2.6\,mag, after
which it is effected by completeness limits. However, despite being
$\sim$2\,mag deeper, the number of sources toward $\cloud$ drops
rapidly above an $(H-K_s)$ of $\sim$1.7\,mag, clearly showing that the
extinction is produced by a sharp increase in dust density as expected
from a dense cloud, as opposed to being produced by several
low-density clouds along the line-of-sight which would be seen as a
much more gradual decrease in the number of red stars.

We used the NACO data and Eq. 2 of \citet{nishiyama2006} to estimate
the distance to the cloud, using the red clump (RC) stars around
$K_s\sim15$ mag.  We assume an absolute magnitude for the RC stars of
$M_K = -1.54$ mag\footnote{ We adopt the \citet{schodel2010} value of
  $M_K = -1.54$ rather than that of $M_k=-1.59$ in
  \citet{nishiyama2006} as we are directly comparing our data to the
  former.}, a population correction $\Delta M_K = -0.07$, and the
extinction law of \citet{schodel2010} for the Galactic Centre
($A_\lambda \propto \lambda^{-2.21}$). The distance modulus is then
given by $\mathrm{DM}=K-M_{K_s}+\Delta M_K$, where $K$ is the
observed, de-reddened $K$ magnitude. We are looking for the distance
to the near side of the cloud, so we used $(H-K_s)=1.7$ mag to
determine the extinction, since this is the color where we start to
lose stars with respect to the Galactic Centre of Schodel et
al. (2010) (see Figure~\ref{fig:extinction}, right-hand panel). Using
the aforementioned extinction law, we derive $A_{K_s}=2.13$ mag for an
effective wavelength of $2.168 \mu$m, which, when applied to the
observed magnitude of the RC stars ($K_{S,obs}=15\pm0.3$ mag), yields
a de-reddened $K_s=12.87\pm0.3$ mag, or $K=12.86\pm0.3$ considering
the difference between $K_s$ and $K$ \citep{nishiyama2006}.  The
distance modulus is then $\mathrm{DM}=14.34\pm0.3$, translating into a
distance of $7.4\pm1.0$\,kpc and independently placing it just
foreground of the Galactic Centre. The greatest source of uncertainty
in this analysis is undoubtedly the extinction law, and we note that a
difference of only 10\% in the exponent of the extinction law in
particular translates into an approximately 10\% change in the derived
distance.

In summary, based on the UKIDSS and NACO data we conclude that we have
undoubtedly detected the extinction caused by $\cloud$, and that this
is a single physical entity (as opposed to multiple clouds separated
by large distances along the line of sight) with a distance consistent
with being close in proximity to the Galactic Centre. Within the
uncertainties, the IR-derived distance to $\cloud$ of $7.4\pm1.0$\,kpc
is consistent with the distance of 8.4\,kpc that would be inferred if
$\cloud$ is part of the ``100\,pc'' ring orbiting the Galactic center,
as proposed by \citet{molinari2011}. The distance to the ``100\,pc
ring'', in turn, is based on the distance to the Galactic center
determined by \citet{reid2009}. As the formal uncertainty in the
distance to the Galactic center is smaller than that from the
IR-derived distance, we adopt a distance to G0.253+0.016 of 8.4\,kpc
for the remainder of the paper. The results of the paper remain valid
if the nearer distance were adopted.

\subsection{Dust temperature, column density, radius and dust mass}
\label{sub:higal_analysis}

HiGAL data of the inner 2$^\circ \times 2^\circ$ of the Galactic
center covering the Central Molecular Zone and $\cloud$ are presented
by \citet{molinari2011}. Because continuum emission at these
wavelengths is typically optically-thin and arises from the dust, we
can use the Herschel data to derive the dust temperature and column
density of $\cloud$.

The dust temperature and gas column density of $\cloud$ were estimated
from a two-step pixel-to-pixel graybody fit to the Hi-GAL data. In the
first coarse-angular-resolution step, graybody fits were calculated
using the 170 - 500 $\mu$m data, scaled to the resolution of the 500
$\mu$m image ($\sim$36$\arcsec$). These results were used as inputs to
the second, high-angular-resolution graybody fits to the 170 - 350
$\mu$m data, which were scaled to the resolution of the 350 $\mu$m
image ($\sim$25$\arcsec$). Both steps used a fixed $\beta$ of 1.75 and
did not include the 70 $\mu$m point due to the high optical depth at
this wavelength. The fitting method is similar to that described in
\citet{bernard2010} and \citet{compiegne2011} but with the diffuse
component of the emission removed \citep[see][for full details on the
  methodology]{battersby2011}.

\begin{figure*}
  \begin{center}
    \begin{tabular}{cc}  
    \includegraphics[width=0.45\textwidth,clip=true]{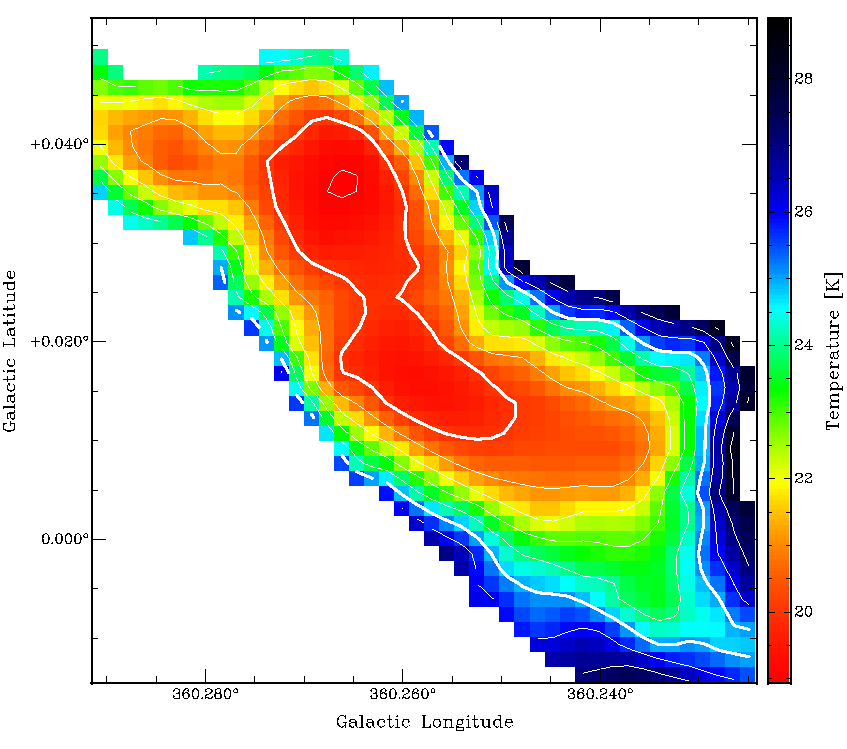} &
    \includegraphics[width=0.45\textwidth,clip=true]{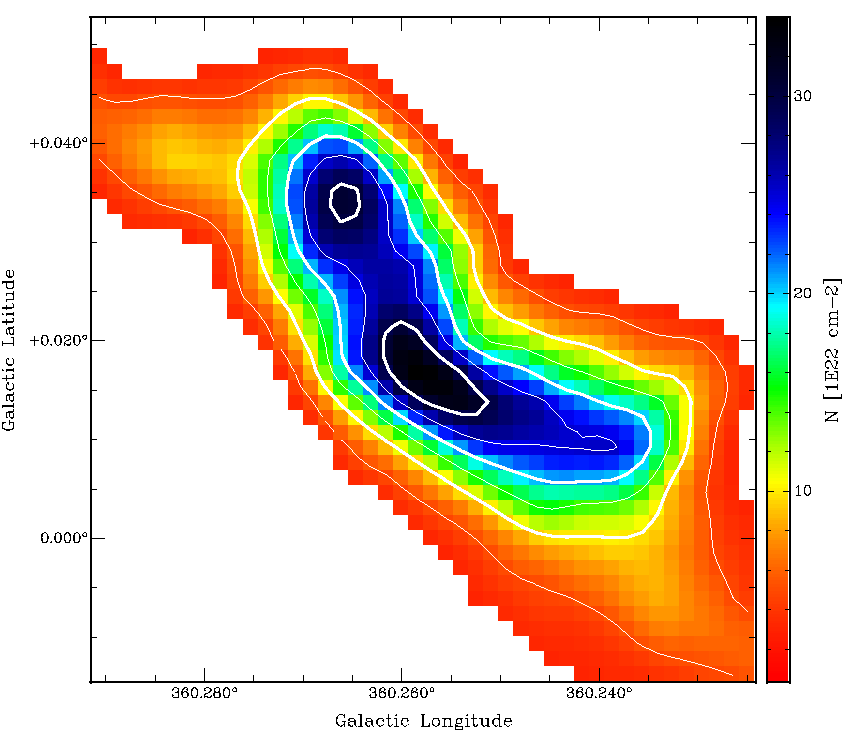} \\
    \end{tabular}
    \caption{Dust temperature [left] and column density [right] maps
      of $\cloud$ derived from the Hi-GAL data. The images are 12\,pc
      on a side for the distance of 8.4\,kpc. The temperature contours
      are 19, 20, ..., 28\,K and the column density contours are 5,
      10, ..., 35\,$\times 10^{22}$\,cm$^{-2}$. The color scale shows
      all pixels above the threshold column density of
      3$\times$10$^{22}$\,cm$^{-2}$ used to define the boundary of
      $\cloud$ (see $\S$~\ref{sub:higal_analysis}). The dust
      temperature overall is low, increasing smoothly from $\sim$19\,K
      at the center to $\sim$27\,K at the edge. There are no obvious
      small pockets of heated dust from any embedded sources. The
      derived external temperature of $\cloud$ is significantly warmer
      ($\gtrsim$35\,K). The column density and dust temperature are
      anticorrelated as expected for an externally-heated, dense
      clump. The derived peak column density is
      $\sim$3.3$\times$10$^{23}$\,cm$^{-2}$ which decreases smoothly
      towards the edge. The average column density in the region above
      the threshold cutoff is
      $\sim$1$\times$10$^{23}$\,cm$^{-2}$. Based on the column density
      maps we derive $\cloud$'s effective radius and mass to be
      2.8\,pc and 1.3$\times$10$^5$\,M$_\odot$, respectively.}
    \label{fig:temp_N_image}
\end{center} 
\end{figure*}

Figure~\ref{fig:temp_N_image} shows the derived dust temperature and
column density maps. We find the dust temperature overall is low,
increasing smoothly from $\sim$19\,K at the center to $\sim$27\,K at
the edge. There are no obvious small pockets of heated dust from any
embedded sources. The derived external temperature of $\cloud$ is
significantly warmer ($\gtrsim$35\,K).

The column density and dust temperature are anticorrelated as expected
for an externally-heated, dense clump. The derived peak column density
is $\sim$3.3$\times$10$^{23}$\,cm$^{-2}$ and decreases smoothly
towards the edge. The cloud area was defined using a column density
threshold of 3$\times$10$^{22}$\,cm$^{-2}$ which covers all the
emission at T$\le 27$\,K. Within this area the average column density
is $\sim$1$\times$10$^{23}$\,cm$^{-2}$. The emission is elongated with
semi-major and semi-minor axes of $1.\arcmin9\times0.\arcmin7$,
corresponding to a physical radius of 4.7$\times$1.7\,pc for a
distance of 8.4\,kpc. Taking the geometric mean, we derive an
effective radius of 2.8\,pc for $\cloud$.

Given the strong dependence of dust emissivity with temperature and to
account for the dust temperature variation across $\cloud$, we
calculate the dust mass directly from the derived column density
rather than from mm continuum emission which assumes a single dust
temperature. The dust mass was calculated by multiplying the column
density of each pixel by the pixel area, assuming a mean molecular
weight of 2.8$\times$M$_{H-atom}$ \citep{kauffmann2008} and
gas-to-dust ratio of 100:1 \citep{hildebrand1983}\footnote{Due to the
  metallicity gradient in the disk \citep{balser2011}, it is possible
  the gas-to-dust ratio in the CMZ may be lower than the canonical
  value of 100:1 commonly adopted in the literature. However, as
  individual metallicity measurements across the CMZ vary from 1 to
  $\sim$4 $\times$ Solar
  \citep{shields_ferland1994,maeda2002,najarro2009} and no independent
  measurements exist for $\cloud$, we opt for the standard gas-to-dust
  ratio of 100:1, while noting the dust-based mass may possibly be
  lower.}, and summing over the pixels in the cloud area. The total
mass is not particularly sensitive to the column density threshold,
varying by $<$10\% for threshold values of
1$-$5$\times$10$^{22}$\,cm$^{-2}$. We derive a dust mass for $\cloud$
of 1.3$\times$10$^5$\,M$_\odot$ and estimate the uncertainty to be of
order 20\%.

\subsection{Molecular line data: linewidth and virial mass}
\label{sub:hops_malt90}

To determine whether $\cloud$ is gravitationally-bound we compared the
dust mass with the virial mass, M$_{\rm vir}$. M$_{\rm vir}$ is
related to the measured radius, R, and linewidth, $\Delta$V, through
M$_{\rm vir} \propto R\, \Delta V^2$, where the constant of
proportionality depends on the geometry and gas density profile
\citep{maclaren1988,bertoldi_mckee1992,dunham2010}. Clearly, small
differences in linewidth can strongly affect the virial mass so it is
important to ensure the measured linewidth reflects the underlying gas
kinematics and has not been affected via other mechanisms
(e.g. different excitation conditions, outflows, shocks, chemistry and
optical-depth effects).

 We used recently available data from the ongoing MALT90 molecular
 line survey \citep{jackson2011,foster2011} to investigate the
 velocity structure of $\cloud$. Detailed analysis and modelling of
 the emission from the many detected molecular line transitions are
 underway and will be presented in a subsequent paper (Rathborne et
 al. in preparation). Here we focus on the global line properties with
 the aim of deriving the most reliable, representative linewidth for
 $\cloud$.

The brightest, optically-thick lines [e.g. HCO$^+$(1-0), HNC(1-0)]
show the velocity structure towards this region is
complicated. Emission is seen from several distinct velocity
components over the spatial extent of $\cloud$ defined in
$\S$~\ref{sub:higal_analysis}. The most prominent of these are: a
component at $\sim$0\,$\kms$ seen at the north-east edge of the
filament; a component at $\sim$35\,$\kms$ covering a similar emission
area as the spatial extent of $\cloud$ defined above; and a component
at $\sim$75\,$\kms$ seen towards the south west edge of the
filament. Given the spatial extent of the 35\,$\kms$ component closely
matches that of the dust emission, we take this component as the
emission from $\cloud$, with the other components being from unrelated
clouds along the line of sight.

The 35$\,\kms$ component shows evidence for shifts in the peak
V$\rm_{lsr}$ with changes in spatial position (ie a velocity
gradient). This means that taking a spectra from an individual spatial
position would not accurately recover the integrated linewidth over
the whole region, as the linewidth-broadening from the velocity
gradient would not be taken into account. To overcome this requires
extracting spectra which are integrated over the spatial extent of
$\cloud$.

\begin{figure*}[!ht]
\centering
  \includegraphics[height=0.75\textwidth,clip=true, angle=-90]{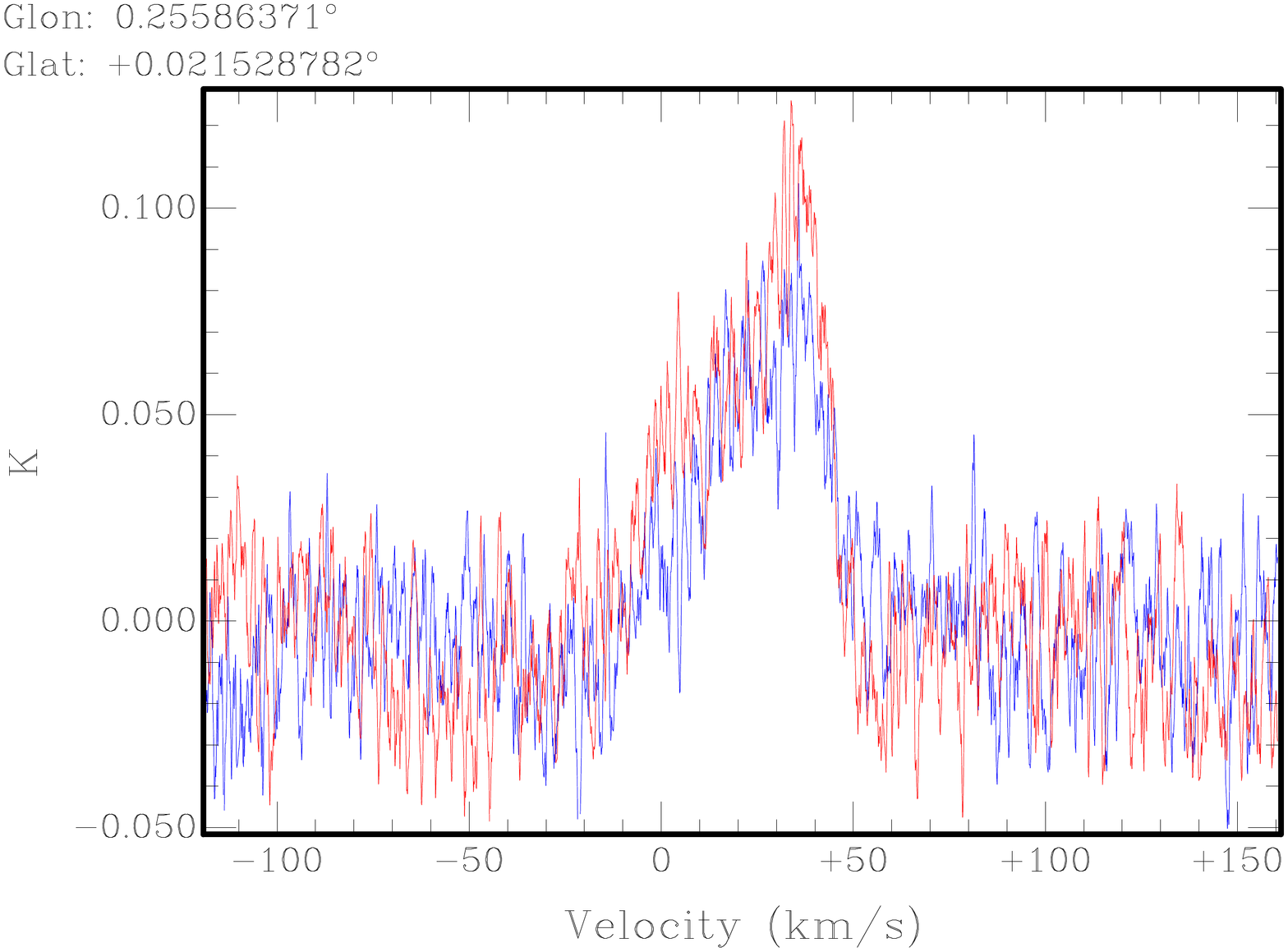} \\
  \includegraphics[height=0.75\textwidth,clip=true, angle=-90]{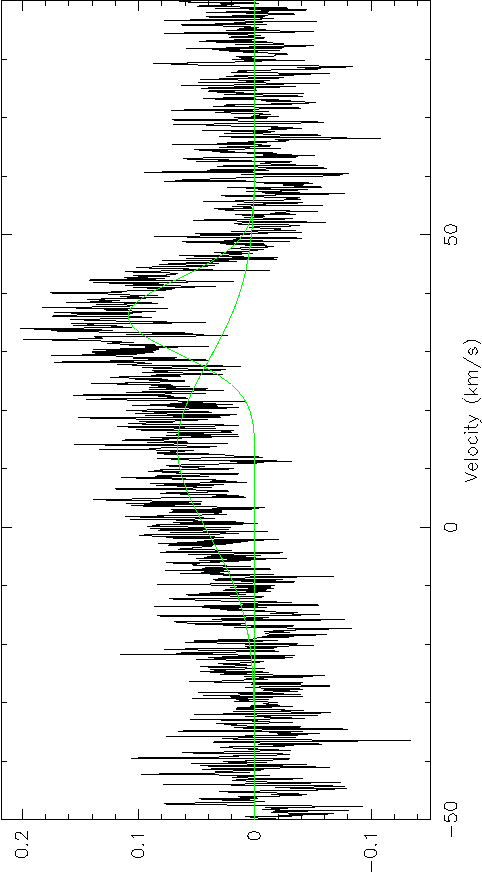} \\
  \caption{[Top] Hanning-smoothed H$^{13}$CO$^+$(1-0) [blue] and
    HN$^{13}$C(1-0) [red] spatially-averaged spectra across $\cloud$
    from the MALT90 survey. Emission from both these transitions is
    expected to be optically-thin so should reliably trace the
    underlying gas kinematics. The line profiles for both transitions
    are similar, with two components: a brighter one at V$\rm_{lsr} =
    35\,\kms$ from $\cloud$ and a weaker one at V$\rm_{lsr} = 0\,\kms$
    from an unrelated cloud along the line of sight. [Bottom] The
    black line shows the same HN$^{13}$C(1-0) line profile as above,
    but without any Hanning smoothing. The green line shows the
    results of a two-component Gaussian fit to this profile, from
    which a line FWHM of $15.1\pm1.0\,\kms$ is derived for the gas
    associated with $\cloud$.}
  \label{malt90_opt_thin_tracers} 
\end{figure*}

We selected bright, optically-thin emission as the most likely
reliable tracer of the underlying gas kinematics. The top panel of
Figure~\ref{malt90_opt_thin_tracers} shows Hanning-smoothed spectra
from the H$^{13}$CO$^+$(1-0) and HN$^{13}$C(1-0) transitions
integrated over the spatial extent of the cloud (as determined in
$\S$~\ref{sub:higal_analysis}). The line profiles for both transitions
are similar, giving confidence that they are robustly tracing the
underlying gas kinematics. Both the V$\rm_{lsr} = 35\,\kms$ component
from $\cloud$ and the V$\rm_{lsr} = 0\,\kms$ component are seen but
the 75\,$\kms$ component is too faint in these optically-thin
transitions to be detected. The bottom panel of
Figure~\ref{malt90_opt_thin_tracers} shows the spatially-averaged
HN$^{13}$C(1-0) emission with no Hanning smoothing. This emission was
fit with a two-component Gaussian profile using the tasks within the
CLASS\footnote{CLASS is part of the GILDAS software environment:
  http://iram.fr/IRAMFR/GILDAS/} software package. To avoid biasing
the fit results we did not constrain any of the parameters but rather
left them all as free variables. From the resulting fits, which are
overlayed on the spectra in the bottom panel of
Figure~\ref{malt90_opt_thin_tracers}, we derive the peak velocity and
FWHM of the gas associated with $\cloud$ to be V$\rm_{lsr}=
36.1\pm0.4\,\kms$ and $\Delta$V$=15.1\pm1.0\,\kms$, respectively. We
adopt an upper limit linewidth of 16\,$\kms$ for $\cloud$ for the
remainder of the paper.


The derived virial mass can vary by an order of magnitude depending on
the assumed geometry and density distribution. Assuming a spherical
source with a density distribution $\propto r^{-2}$ gives a virial
mass estimate of 9$\times$10$^4$\,M$_\odot$ \citep[][Eq
  3]{maclaren1988}. However, more realistically, for a
centrally-condensed cloud elongated with axis ratio 3:1 and density
$\propto r^{-1.8}$, the virial mass would be
4$\times$10$^5$\,M$_\odot$
\citep{bertoldi_mckee1992,dunham2010}. Despite the assumptions and
systematic uncertainties, the dust and virial mass estimates agree to
within a factor of a few. Thus it is likely $\cloud$ is close to
virial equilibrium.

\section{Discussion}

The global properties of $\cloud$ derived in $\S$~\ref{sec:results}
are summarized in Table~\ref{tab:global_properties}. Based on these
results, $\cloud$ is a strong candidate precursor of an Arches-like
YMC. The only other two similar Galactic YMC-forming clouds that have
been studied to date are Sgr B2 and W49A
\citep[e.g.][]{goldsmith1990,alves2003}. However, these are both much
more evolved, with massive cluster formation well underway. The
powerful feedback from ongoing star formation has strongly affected
the cloud structure in these regions, so observations of the gas do
not probe the initial conditions prior to the onset of star formation.

We speculate that $\cloud$ may be unique in representing the initial
conditions of a Galactic molecular cloud on the verge of forming a
YMC. As such, its detailed study can reveal important clues about
massive cluster formation and help test theoretical models. Current
theories suggest that the early evolution of stellar clusters depends
crucially on the spatial and kinematic distribution of stars
\citep[e.g.][]{goodwin_bastian2006,allison2010}. However, while the
distribution of emergent stellar populations as ``hierarchical'' or
``centrally-condensed'' may be straightforward to quantify when
analyzing stars, quantifying the small-scale distribution of the gas
in a similar way is complicated \citep[see e.g.][]{lomax2011}. Because
gas can not easily be counted in discreet, observationally-defined
units, the results of algorithms attempting to quantify gas
substructure are notoriously subjective to the values of user-defined
inputs \citep[see e.g.][]{pineda2009}.

Regardless of the exact details of the formation mechanism, the
properties of the emergent stellar population must be directly related
to the initial global gas properties. Assuming a reasonable star
formation efficiency (SFE) \citep[e.g. 30\% for gas
  $\geq10^4\,$cm$^{-3}$;][]{alves2007}, $\cloud$ may form a cluster
with stellar mass and density similar to that of other YMCs like the
Arches. Whether or not such a cluster remains gravitationally-bound
depends on the SFE and environmental conditions \citep[tidal shear,
  interaction with GMCs etc,
  e.g.][]{kruijssen2011,elmegreen_hunter2010}. However, it is clear
that $\cloud$ has enough mass to form an Arches-like cluster without
accreting any additional gas or stars from outside the present-day
observed boundary.

With only a single candidate YMC precursor it is not possible to
determine how representative $\cloud$ is of all YMC progenitors.  It
is clearly desirable to search for other proto-YMC molecular clouds,
especially within the Galaxy. In the last decade much observational
effort has been directed towards searching for, and characterizing the
physical properties of molecular clouds which are likely the
progenitors of stellar clusters. As a result, a large fraction of the
gas in the Galaxy available to form YMCs has been surveyed in dust
continuum and dense molecular gas tracers.

\begin{figure*}
 \begin{center}
 	\includegraphics[width=1.0\textwidth,angle=0]{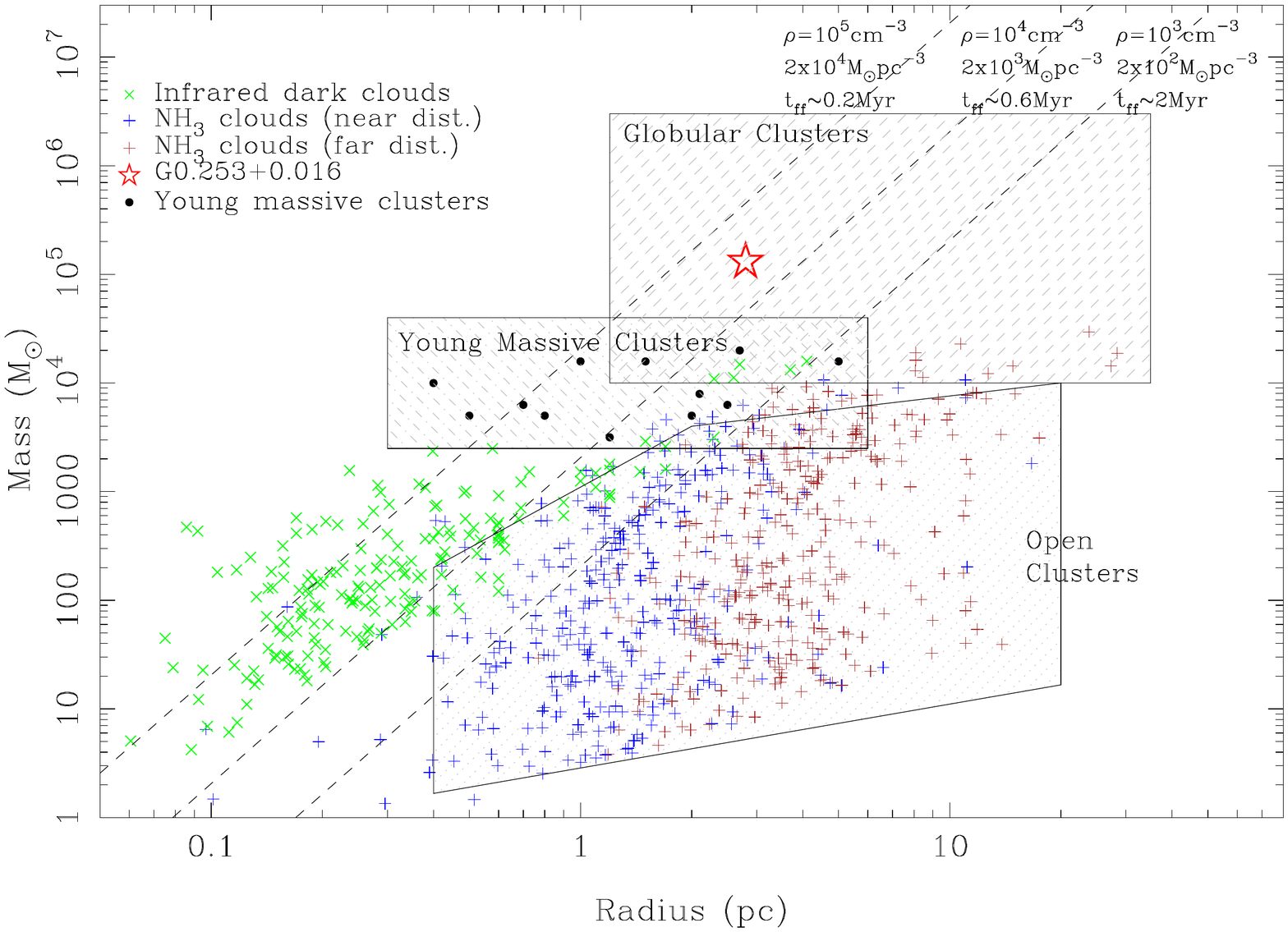}
 	\caption{Radius versus mass for Galactic dense,
          cluster-forming molecular clouds. Plus symbols show ammonia
          clouds detected in HOPS \citep{walsh2011} (blue/brown denote
          an assumed near/far kinematic distance, respectively). Green
          crosses show infrared dark clouds (IRDCs) from the survey of
          \citet{rathborne2006irdc}. The hatched rectangles show the
          mass-radius range of different stellar clusters
          \citep[][]{portegieszwart2010}. The black dots show Galactic
          young massive clusters. With the exception of a few clouds
          which may form small YMCs, assuming a reasonable star
          formation efficiency, most of the observed molecular clouds
          seem destined to form open clusters. $\cloud$ is marked with
          a red star and clearly stands out as unique. It has a mass
          and radius that would be expected of a molecular cloud
          progenitor of a large YMC or a globular cluster. The dashed
          lines show constant density and free-fall time.}
    	 \label{fig:radius_mass}
 \end{center} 
 \end{figure*}

Figure~\ref{fig:radius_mass} summarizes the results from two of these
surveys and places $\cloud$ in the context of Galactic molecular
clouds and stellar clusters. Plus symbols show ammonia clumps detected
in HOPS \citep{walsh2011}. The critical density of ammonia is close to
the volume density threshold of $\sim$10$^4$\,cm$^{-3}$ proposed by
\citet{lada2010} as the threshold required to form stars. As such HOPS
is ideal for identifying and characterizing the properties of star
forming gas within the Galaxy.  Green crosses show IRDCs from the
survey of \citet{rathborne2006irdc}. IRDCs are thought to be
precursors of high mass stars and clusters. The region of the
mass-radius plane covered by the HOPS clouds and IRDCs is similar to
that of clouds detected in other surveys, both targeted \citep[e.g. as
  compiled by][]{kauffmann_pillai2010} and blind, large-area surveys
(BGPS -- \citet{aguirre2011}; ATLASGAL -- \citet{schuller2009},
Tachenberg et al submitted and Beuther et al. priv. comm). The hatched
rectangles show the mass-radius range of different stellar clusters
\citep[][]{portegieszwart2010} and the black dots show Galactic
YMCs. With the exception of a few clouds which may form small YMCs,
assuming a reasonable star formation efficiency, most of the observed
molecular clouds seem destined to form open clusters. $\cloud$ is
marked with a red star and clearly stands out. It is significantly
more massive for its size compared to the other Galactic molecular
clouds. It is potentially the only object with an observed gas mass
and density which may be able to directly form a cluster with stellar
mass and densities in the YMC/globular cluster regime.

The potential uniqueness of $\cloud$ has profound implications for YMC
formation in the Galaxy.  We posit three scenarios. If all YMCs form
from molecular clouds with similar properties to $\cloud$, either i)
$\cloud$ is unique in the Galaxy, or, ii) surveys have missed the
other similar clouds. Alternatively, iii) if the properties of
$\cloud$ are not representative of other molecular clouds destined to
form YMCs, there may be precursors to YMCs in the Galaxy which have
already been observed but not identified as YMC precursors.

To distinguish between these scenarios it is important to predict how
many proto-YMC molecular clouds are likely to exist at any given time
in the Galaxy.  One way to estimate this is following the analytical
approach of \citet{gieles2009}. Assuming a star formation rate of 1
M$_\odot$/yr with 10\% of stars forming in clusters and a cluster mass
function index of $-2$, we would expect $\sim$10 clusters of
$>$10$^4$\,M$_\odot$ with ages $<$10\,Myr. This estimate is consistent
with the number of YMCs currently known in the Galaxy. If the number
of Galactic YMCs is constant, assuming it takes of order a Myr to
evolve from precursor to forming cluster we would expect from zero to
a few precursors at any given time. An alternative method to estimate
the expected number of proto-YMC molecular clouds in the Galaxy is
extrapolating the observed embedded cluster mass function (ECMF),
which, for embedded clusters up to $10^3$ M$_\odot$ within 2 kpc of
the Sun, is a power-law with index $-2$
\citep{lada_lada2003}. Assuming this represents the Galactic ECMF
\citep[see][]{gieles2009}, one can scale it to the total volume of the
Galactic disk to estimate the number of embedded clusters expected in
the Galaxy. Extrapolating this to higher cluster masses, we expect
$\sim 25$ young clusters with masses between $10^4$ and $10^5$
M$_\odot$. Assuming a cluster lifetime of 10 Myr we predict $\sim 2.5$
proto-YMCs per Myr in the Galaxy. Both of these approaches suggest
there may be other Galactic YMC progenitors. While these estimates are
approximate, they predict that only a small number of proto-YMCs
should exist in the Galaxy at any given time. However, the
uncertainties are sufficiently large that it is not possible to use
these estimates to rule out $\cloud$ as being unique. Therefore, it is
currently not possible to distinguish between scenarios i) and ii).

Returning to scenario iii), there are many molecular cloud complexes
known with sufficient mass to form YMCs but their present-day gas
density is much lower than the expected final YMC stellar
density. Examples of these are seen clearly in
Figure~\ref{fig:radius_mass} as clouds with mass $>10^4$\,M$_\odot$
and radii $>\sim8$\,pc. Such clouds seem destined to form massive, but
unbound, associations. However, it may be possible to form the
required YMC stellar densities over time from the convergence of
molecular gas flows on large scales and subsequent global
gravitational collapse. Indeed, such large-scale gravitational
collapse has been directly observed towards a number of massive
protoclusters (e.g. W33A \citep{galvan-madrid2010}, DR21
\citep{schneider2010}, G10.6 \citep{keto1988,baobab2010}, G8.67
\citep{longmore2011}) and is predicted by numerical simulations
\citep{smith2009,klessen_hennebelle2010}. In this scenario, star
formation would proceed over several crossing times, leading to a
large age spread in cluster members.  In this regard, it is
interesting to note that an age spread of $\ge 10$\,Myr among cluster
members are reported in several YMCs \citep[NGC~3603, NGC~346,
  30~Doradus --][]{beccari2010,deMarchi2011A,deMarchi2011B}. However,
there is some debate in the literature over the uncertainty in the age
determination of cluster members using this method and whether or not
these data are also consistent with age spreads of $\le$3\,Myr
\citep{tobin2009,baraffe2009,littlefair2011,jeffries2011}. Detailed
studies of the large-scale gas motions in the youngest and most
massive molecular clouds are needed to determine whether the
gas/stellar density will increase over time to form YMCs as opposed to
more diffuse and gravitationally-unbound OB associations as would be
predicted by their current global gas density.

With this scenario in mind, it is interesting to note that large-scale
emission from shocked-gas tracers is detected towards
$\cloud$. Combined with gas temperatures of 80\,K \citep[][Rathborne
  et al. in prep]{lis2001}, which is much warmer than the dust
temperature, this suggests $\cloud$ may have formed from a
cloud-cloud collision or convergence of large scale flows. As the gas
has not yet had the chance to heat up the dust, this must have
happened recently. If this is the case, before the postulated
collision/convergence, the gas properties of $\cloud$ may have been
much more similar to the typical Galactic molecular cloud population
and thus may not have stood out as extreme in
Figure~\ref{fig:radius_mass}.

Given the location of $\cloud$ near to the Galactic center, it is
tempting to invoke the extreme conditions at the Galactic center
(e.g. interstellar radiation field enhanced by 10$^3$, external
pressure P/k$\sim$10$^8$, strong magnetic fields, gamma rays) to
justify why it may be unique. In this regard it is interesting to note
that $\cloud$ lies at a projected distance of only 18\,pc from the
Arches and Quintuplet clusters. While it is not possible to directly
link the formation of the Arches with $\cloud$, it seems plausible
that the Arches could have formed at its present distance from the
Galactic center. On larger scales $\cloud$ appears to be part of a
10$^6$\,M$_\odot$ filament of interconnected clumps of which $\cloud$
is the most massive \citep{lis2001}. This filament is itself thought
to be part of the 3$\times$10$^7$\,M$_\odot$, 100\,pc ring identified
by \citet{molinari2011}.  The 100\,pc ring is orbiting the Galactic
center at $\sim$80\,$\kms$ so a single orbit would take
$\sim$3.6\,Myr. Given the free-fall time for $\cloud$ is $<$1\,Myr
(see Table~\ref{tab:global_properties}), it seems unlikely $\cloud$
would survive a whole orbit without additional support -- not
implausible given the extreme conditions at the Galactic center. If
such additional support could be maintained against the rapid
turbulent energetic decay, this may explain why the cloud has remained
dense and starless.

Alternatively, the fate of the $\cloud$ may be more intimately linked
to the dynamics of the 100\,pc ring. Gas is thought to enter the ring
at the intersection points of the innermost stable X1 orbits and the
X2 orbits. These intersection or ``crashing'' points coincide with Sgr
B2 and Sgr C -- the only places in the 100\,pc ring with prodigious
star formation activity. If the gas comprising $\cloud$ were to have
entered the 100\,pc ring at Sgr C it would have joined only a small
fraction of an orbit time ago. In around a free-fall time, the whole
10$^6$\,M$_\odot$ filament in which $\cloud$ is located will have
reached the position of Sgr B2. As such, this larger filament could be
in the process of forming a 10$^6$\,M$_\odot$, Sgr~B2-like complex.

However, while it is tempting to invoke the extreme conditions at the
Galactic center to justify why $\cloud$ may be unique, several of the
most massive YMCs in the Galaxy are not located near the Galactic
center suggesting the extreme environmental conditions are not a
necessary condition to form a YMC.

\section{Conclusions}

We make use of existing multi-wavelength data including recently
available far-IR continuum (Herschel/Hi-GAL) and mm spectral line
(HOPS and MALT90) data and present new, deep, multiple-filter, near-IR
(VLT/NACO) observations to study the infrared-dark cloud
$\cloud$. From these data we have derived the global properties of
$\cloud$. It has a high mass (M$_{\rm dust}
\sim1.3\times10^5$\,M$_\odot$), low dust temperature ($\sim$23\,K),
small radius ($\sim$2.8\,pc), high volume density
(7.3$\times10^4$\,cm$^{-3}$) and high column density
(3.5$\times10^{23}$\,cm$^{-2}$). It is close to virial equilibrium
with almost no signs of active star formation. As such it appears to
be a prime candidate for the initial conditions of a molecular cloud
destined to form an Arches-like YMC.

Comparing the properties of $\cloud$ to other Galactic dense molecular
clouds shows it to be extreme. We discuss implications of this for the
formation of massive protoclusters and posit three scenarios. If all
YMCs form from molecular clouds with similar properties to $\cloud$,
either i) $\cloud$ is unique in the Galaxy, or, ii) surveys have
missed the other similar clouds. Currently it is not possible to
distinguish between scenarios i) and ii). Nevertheless, clouds with
properties like $\cloud$ must be very rare in the Galaxy making
$\cloud$ extremely important for testing massive cluster formation
models. Alternatively, in scenario iii) the properties of $\cloud$ do
not represent those of other molecular clouds destined to form
YMCs. We note there are many molecular clouds in the Galaxy with
sufficient mass to form YMCs but their present-day gas density is much
lower than the expected YMC stellar densities. However, it may be
possible to form the required stellar densities over time from the
convergence of molecular gas flows on large scales and subsequent
global gravitational collapse, as observed towards several lower-mass
protoclusters. Detailed studies of the large-scale gas motions in the
youngest and most massive molecular clouds are needed to determine
whether the gas/stellar density will increase over time to form YMCs
as opposed to more diffuse and gravitationally-unbound OB associations
as would be predicted by their current global gas density.

In future work we will study the internal structure and kinematics of
$\cloud$ in detail to investigate its recent dynamical history, test
the plausibility that it formed from cloud-cloud collisions or
converging large-scale flows, and directly test models of massive
protocluster formation.

\section{Acknowledgments}

We thank the anonymous referee for constructive comments that improved
the manuscript. SNL would like to thank Jens Kauffmann, Thushara
Pillai, Qizhou Zhang, Henrik Beuther, Peter Schilke and Adam Ginsburg
for useful discussions. This research made use of the NASA
Astrophysical Data System. The research leading to these results has
received funding from the European Community's Seventh Framework
Programme (/FP7/2007-2013/) under grant agreement No 229517.


\begin{table*}
\begin{center}
  \caption{Global properties of \cloud. The columns show mass (M),
    distance (D), radius (R), dust temperature (T$_{\rm dust}$),
    linewidth ($\Delta$V), volume density ($\rho$), column density
    (N$_{H_2}$), cloud-crossing time ($t_{cc}$), sound-crossing time
    ($t_{sc}$) and free-fall time ($t_{ff}$). }
  \begin{tabular}{cccccccccccc}
    \hline
    M          & D     & R     & T$_{\rm dust}$ & $\Delta$V & $\rho$     & N$_{H_2}$  & $t_{cc}$ & $t_{sc}$ & $t_{ff}$ \\
    (M$_\odot$) & (kpc) & (pc) & (K)          & (km/s)    &(cm$^{-3}$) & (cm$^{-2}$) & (Myr)   & (Myr)   & (Myr)     \\ \hline
    1.3E5      & 8.4   & 2.8   & 19-27        & 16        & 7.3E4      & 3.5E23    &  0.17   & 8        &  0.74    \\ \hline\hline

  \end{tabular}
  \label{tab:global_properties}
\end{center}
\end{table*}


\bibliography{brick_accepted}
\bibliographystyle{apj}


\end{document}